\documentclass[amssymb,aps,epsf,eqsecnum,nofootinbib,showpacs]{revtex4}
\usepackage{graphicx}

\usepackage{makeidx}
\makeindex

\def\bq{\begin{quote}}
\def\eq{\end{quote}}
\def\bqu{\begin{quotation}}
\def\equ{\end{quotation}}
\def\n{\nonumber}
\def\bv{\begin{verse}}
\def\ev{\end{verse}}
\def\ben{\begin{enumerate}}
\def\een{\end{enumerate}}
\def\bi{\begin{itemize}}
\def\ei{\end{itemize}}

\def\bev{\begin{verbatim}}
\def\bw{\begin{widetext}}
\def\ew{\end{widetext}}
\def\eev{\end{verbatim}}
\def\bc{\begin{center}}
\def\ec{\end{center}}
\def\h{\huge}

\def\tbf{\textbf}

\def\tit{\textit}

\def\ra{\rightarrow}

\def\np{\newpage}
\def\s{\section}
\def\ss{\subsection}
\def\sss{\subsubsection}

\def\bb{\bibitem}
\def\ra{\rightarrow}
\def\lora{\longrightarrow}

\def\tg{\tilde{\Gamma}}

\newcommand{\be}{\begin{equation}}
\newcommand{\ee}{\end{equation}}
\def\bqn{\begin{eqnarray}}
\def\eqn{\end{eqnarray}}
\def\n{\nonumber}
\def\p{\partial}
\def\f{\frac}

\def\bmat{\begin{matrix}}
\def\emat{\end{matrix}}

\begin{document}
\title{\h Presymplectic Geometry and the Problem of Time. Part 1\\}

\author{Vasudev~Shyam}
\email{vasudev@cfrce.com}
\author{ B~S~Ramachandra}
\email{bsr@cfrce.com}
\affiliation{Centre for Fundamental Research and Creative Education,\\
Bangalore, India}

\date{\today}
\begin{abstract}
An effective mathematical framework based on Presymplectic Geometry for dealing with the "phase space picture" of timeless dynamics in General Relativity is presented. In General Relativity, the presence of the scalar Hamiltonian constraint which vanishes leads to the problem of time, which can be solved, up to an extent by adopting a timeless formalism. This has been done by Carlo Rovelli and Julian Barbour et al. In this paper we present a phase space reformulation of Barbour's theory. The Presymplectic dynamics of general totally constrained, reparametrization invariant theories is developed, then applied to Jacobi mechanics, relational particle mechanics and the dynamics of the free relativistic particle.

\end{abstract}


\maketitle


\s{Introduction}
The problem of time  has been one of the most extensively investigated topics in quantum gravity.  Due to its key role, its implications have been studied from philosophical, conceptual, structural and mathematical points of view. Not only has the foundational aspects of canditate quantum theories of gravity been much clarified but also the possible meaning of time itself, especially at a fundamental level. Of the several approaches that have come to the forefront of curent research, two are markedly similar. One, pioneered by Rovelli, seeks to "forget time" conceptually. The other, initiated and developed by Barbour, does away with time from the very outset. This latter approach has recently been the topic of extensive investigation by a host of researchers. However, a unified mathematical formalism that embodies these ideas would be instructive and desirable. The presymplectic framework presented in this paper seeks to do exactly this. In the first of a two-part paper we develop this formalism in the context of general totally constrained, reparametrization invariant theories. We then apply it to Jacobi mechanics, relational particle mechanics and the free relativistic particle. The application of this to the case of gravity is treated in Part 2 of this paper. \\

This Paper is organized as follows. In Section II we very briefly state the problem of time. In Section III, we discuss the Jacobi-Barbour-Bertotti theory. In Section IV we make some remarks on Ephemeris time and its role. In Section V we take a look at Rovelli's theory. In Section VI we discuss the violation of Dirac's theorem for completely constrained systems. In Section VII we treat phase space as a presymplectic manifold. Section VIII presents the presymplectic formalism. Section IX carries some concluding remarks.

\s{The problem of time in quantum gravity}

The problem of time in quantum gravity is perhaps as old as the subject itself. It was noted first by Bryce de Witt in his paper "The Quantum Theory Of Gravity"\cite{dw67} in which he presents the constraint equations for canonical Quantum General Relativity, 
\bqn \pi \Psi=0 \\
\pi^{i} \Psi=0\\
\mathcal{H} \Psi=0\\
\chi^{i} \Psi=0
\eqn
where the first two constraints correspond to the vanishing of momenta associated to the Lapse and the Shift, while the other two are the scalar Hamiltonian constraint and the vector Diffeomorphism constraint respectively. He then notes that in such a setting, if we were to consider the wave function $\Psi$ as a function $\Psi(x,x^{0})$ of the spatial coordinates and time, we see that the expectation value of the field operator, let's say, the momentum conjugate, undergoes something peculiar: $\Psi^{\dagger}\pi^{ij}(x,0)\Psi=\Psi^{\dagger}\pi^{ij}(x,x^{0})\Psi$\\
 This leads him to conclude
\bq ".....Since the statistical results of any set of observations are ultimately expressible in terms of expectation values, one therefore comes to the conclusion that nothing ever takes place in quantum gravitydynamics, that the quantum theory can never yield anything but a static picture of the world."\eq 

Thus the Wheeler de Witt equation, in the standard interpretation leads to the "frozen formalism" which apparently tells that nothing should happen in Canonical Quantum Gravity. So how can one attain at least a partial solution to this problem? In order to answer this, we must first note that the problem arises fundamentally due to the fact that the Hamiltonian is but a sum of constraints, and that they in themselves saturate the theory. Due to the absence of the unconstrained Hamiltonian, we are left with no operator which can generate infinitesimal time translations of the system.\\
Now here's the question we shall attempt to answer in the following part of this paper and the next: How can we extract dynamics from a system that is totally constrained? We shall adopt the "No Time" interpretation of Barbour et. al. for both the present and the forthcoming models including canonical general relativity.

\s{Jacobi-Barbour-Bertotti theory}
We shall first attempt to understand what it means to eliminate "time" from the action. The essence of doing so would be to replace time by an unphysical evolution parameter, with respect to which we can differentiate coordinates to attain velocities, which can indeed exist with the notion of an absolute or even a clock time, and the action should really be invariant under an arbitrary change of this parameter, i.e be reparameterization invariant. Thus eliminating time really amounts to getting rid of a preferred time scale, i.e. the system should refer to no external clock. As we shall see in the coming chapters, in doing so, the system finds a way to be its own clock! Now if we considered $\lambda$ to be the evolution parameter, this condition implies that $\lambda \rightarrow f(\lambda)$ should leave the action invariant. As we shall see, the Action for an ordinary nonrelativistic particle dynamics just wont do. This is because,
\be S'=\int d\lambda f'(\lambda)\left(\frac{1}{f'(\lambda)^{2}}T-U\right) \ee
where, all the primes and velocities here denote derivatives with respect to $\lambda$). \\
As it is clearly evident, the Action is not invariant under such a transformation, largely due to the dependence of the Lagrangian on the square of the first time derivative of the position. So in order to attain a reparameterization invariant Action, we need to find an Action that is linear in velocities, so that we can perform a Routhian reduction by letting the Newtonian time 't' be a function of $\lambda$ and eliminating it. In doing so we obtain the Action
\be S_{JBB}=\int d\lambda 2\sqrt{(T(E-U))} \ee
which is the Jacobi Barbour Berotti action that obeys the Jacobi geodesic variational principle, which tells us that the geodesics on configuration space extremize the Action.
\s{Ephemeris Time}
As mentioned earlier, for the system we just considered, the removal of time scale via reparameterization invariance of the Action would lead to an interesting result. The system becomes its own relational "clock", i.e. the notion of time is derived from dynamics. In order to exhibit this, we shall use a rather pedestrian derivation of the "Ephemeris Time".
We consider
\bqn T=E-U \\
=> \frac{1}{2}m \eta_{ab}\frac{dq^{a}}{dt}\frac{dq^{b}}{dt}=E-U\\
=> dt=\sqrt{\frac{m \eta_{ab}dq^{a}dq^{b}}{2(E-U)}}
\eqn
where, the $\eta$ is the Euclidean metric. \\
What we have here, in itself is a remarkable result, it is as though an increment of time is the "mass weighted hypotenuse" of the total dynamical increment in the system. Now if we make our earlier assumption that velocities should be defined over an arbitrary evolution parameter, we see that the above expression becomes,
\bqn dt=\sqrt{\frac{T}{E-U}}d\lambda \\
=> t_{y}-t_{x}=\int_{x}^{y} \sqrt{\frac{T}{E-U}}d\lambda
\eqn
Thus it is evident that the Ephemeris gives us a \tit{metric-trans-temporal} notion of identity between two subsequent configurations, and time is now completely emergent from the dynamics. We shall not take a look at Rovelli's theory.
\s{Rovelli's theory}
Here we shall breifly discuss Carlo Rovelli's timeless theory. We shal be using the same mathematical framework developed by Rovelli et. al. but our formalism differs subtly in interpretation and mostly in application, as we use this formalism via the Barbour interpretation and not the evolving constants of motion. In this theory, the Phase space $T^{*}\mathcal{C}\ni (q_{a},p^{a})$ is a 2n dimensional manifold that admits a k dimensional surface $\Sigma$ and carries the canonical one form
\be \tilde{\theta}=p_{a}dq^{a}\ee
and $\tilde{\theta}|_{\Sigma}=\theta$, now, the dynamics is governed by the equation
\be d\theta(X)=0 \ee
 where, $X$ belongs to the $2n-k$ dimensional kernel of $d\theta$. Now the motions are integral surfaces of this equation. The Hamilton-Jacobi formalism of this theory is as follows. The Hamilton-Jacobi equation is given by
\be H\left(q^{a}, \frac{\partial S(q^{a})}{\partial q^{a}}\right)=0 \ee
where, $S(q^{a})$ is Hamilton's principal function. This admits a one parameter family of solutions $S(q^{a},Q^{i})$ parameterized by $n-k$ constants of motion $Q^{i}$ and the equation for the "motions" of the system is given by,
\be f^{i}(q^{a},Q^{i},P_{i})=\frac{\partial S}{\partial Q^{i}}-P_{i}=0 \ee
for $n-k$ constants $P_{i}$. $Q, P$ coodinatize the $2(n-k)$dimensional phase space.
\s{Violation of Dirac's theorem}
It is instructive to note here the invalidity of Dirac's theorem regarding first class constraints for completely constrained Hamiltonian systems. 
\ss{Systems with first class constraints}
We begin by considering constrained Hamiltonian systems with first class constraints.
The Hamiltonian is given by
\be 
H_{T} = H + v_{a}\phi_{a}
\ee 
where the $ v_{a}$ are arbitrary Lagrange multipliers and the $\phi_{a}$ are the primary first class constraint functions. For some function $f$ which is a dynamical variable, an infinitesimal increment in time $\delta t$ is given by
\bqn \delta f = \delta t \{f, H_{T}\} \\
= \delta t (\{f, H\} + v_{a}\{f, \phi_{a}\})
\eqn
Due to the arbitrariness in the $v_{a}$, we can alternatively choose $v^{\prime}_{a}$ and the difference
\be
\Delta f(\delta t) = \epsilon_{a}\{f, \phi_{a}\}
\ee
where,
\be
\epsilon_{a} = \delta t(v_{a} - v^{\prime}_{a})
\ee
is very small but due to the fact that such changes of $v_{a}$ should not lead to physically inequivalent motions, it is concluded that $\phi_{a}$ generates unphysical gauge tansformations. We must note that Dirac assumes that $\delta t$ is the same for both $v_{a}$ and $v^{\prime}_{a}$. This is the crux of the matter as it tells us how absolute time influences this interpretation.
\ss{Repametrization invariant models}
Here the Hamiltonian is a constraint
\be
H_{T} = v_{a}\phi_{a}
\ee
an example of which is the Barbour-Bertotti theory where
\be
H = N (\f{p_{a}p^{a}}{2m} + V - E)
\ee
We note that in this model there is no preferred notion of time. Increments of time are determined by the motion of the system via ephemeris time
\be
\delta t = N \delta \lambda
\ee

where
\be
N = \sqrt{\f{m\delta q_{a} \delta q^{a}}{2(E - V)}}
\ee
that has a corresponding expression in phase space
\be
N = \sqrt{\f{ \eta^{ab} p_{a} p_{b}}{2m(E - V)}}
\ee
For the dynamical evolution of some $f = f(q_{a}p_{a})$ we have
\bqn
\delta f(q_{a}p_{a}) &=& N\delta \lambda\{f, H\} \\
&=& N\delta \lambda \{f, \f{ \eta^{ab} p_{a} p_{b}}{2m}+U-E\}
\eqn
which is a physical motion!
Now in analogy to what we previously did, we have
\be
\Delta f = \{f, H\}(N\delta \lambda)
\ee
And by changing $N$ we inadvertently change $\lambda$ because ephemeris time determines the parametrization. So if we are to consider changing $N$ we would change $\delta \lambda$, or vice versa, in such a manner that there is no diference and now there is no ambiguity in the evolution of $f$. We see this as a clear indication to choose a phase space with no preferred \tit{apriori} assumed parametrization and the presymplectic manifold is just what we need.
\s{Phase space as a presymplectic manifold}
We recall from the symplectic formulation that the Legendre transformation is the fibre derivative that takes us from the configuration to the phase space.\\
For some Lagrangian $L(q, \dot(q)) \quad \epsilon \quad \cal{C}$ the fibre derivative is a map $FL $:$ T\cal{C} \ra T^{*} \cal{C}$
In the usual unconstrained case

\be
P \circ FL = \f{\p L}{\p\dot{q}}
\ee
exists for all $\dot{q}'s$ and so
\be
H \circ FL = p\dot{q} - L
\ee
is unconstrained. Thus we attain a symplectic form on phase space which is strongly non-degenerate. i.e there exists an $X_{H}$  $\epsilon$  $TT^{*}\cal{C}$ such that
\be
\iota_{X_{H}}\omega = dH
\ee
This is due to $L_{X}\omega$=$0$ that implies $\iota_{X}d\omega$+$d\iota_{X}\omega$=$0$  since$T^{*}\cal{C}$ is simply connected. This gives $L_{X}\omega$=$dH$ and $X$=$X_{H}$. Thus, in such a system where the fibre derivative is a total diffeomorphism, the map
\be
\flat : TT^{*}\cal{C} \ra T^{*}T^{*}\cal{C}
\ee
is bijective. Thus the symplectic two-form is strongly non-degenerate. We shall refer to this as the symplectic case. Henceforth for simplicity we shall write $T^{*}\cal{C}$ $\equiv$ $\Gamma$. In the completely constrained case given in our example above
\be
H = v_{a}\phi_{a}
\ee
the fibre derivative $H \circ FL = 0$ giving a singular Lagrangian. \\
Thus the condition that the symplectic form be strongly non-degenerate is to be relaxed and so we choose a weakly non-degenerate presymplectic form. The question that now arises is as to where and how exactly does the map $\flat$ act as a bijection (locally)? The answer is that it acts on the foliations of phase space which are the constraint submanifolds and it does so by means of the presymplectic algorithm.
\s{Presymplectic Formalism}
On the total phase space given by $(\Gamma, \omega)$ the symplectic case gives
\be
\iota_{X}\omega = dH
\ee 
but here, in the first class case (it can easily be generalized into any number of constraints, but we choose first class for simplicity) there exists a constraint submanifold $\tilde{\Gamma} \subset \Gamma$ on which the two form is strongly non-degenerate. Thus we choose an inclusion mapping
\be
j_{1} : \tilde{\Gamma} \ra \Gamma
\ee
for which
\be
(\iota_{X}\omega - dH)\circ j_{1} = 0
\ee
In general, the presymplectic algorithm generates a string of submanifolds,
\be
\stackrel{j_{3}}{\lora} \tilde{\Gamma_{3}} \stackrel{j_{2}}{\lora} \tilde{\Gamma_{2}} \stackrel{j_{1}}{\lora}\tilde{\Gamma} \stackrel{j}{\lora} \Gamma
\ee
and this ends up in three cases-\\
Case 1: We hit an $\tilde{\Gamma_{l}}$ such that $\tilde{\Gamma_{l}} = \Phi$. This plainly means that the manifold does not describe a dynamical system.\\
Case 2: We get a $\tg_{l}$ such that dim$(\tg_{l}) = 0$, $\tg_{l}\neq \Phi$. This means that the system is static. i.e $X = 0$\\
Case 3: There exists an $l$ such that $\tg_{l} = \tg_{l+1}$. Thus we have a whole set of submanifolds with $dim(\tg_{l}) \neq 0$. This is the only case that gives us physically relevant information. And from the equivalence of this algorithm and the Dirac-Bergmann theory of constraints, this is the submanifold on which all the constraints are satisfied. Thus we can compound
\be
j_{n}\circ j_{n-1}\circ...\circ j_{3}\circ j_{2}\circ j = \Pi
\ee
and so,
\be
\Pi : \tg_{n} \lora \Gamma
\ee
Putting it all together we find that for completely constrained systems
\be
[\iota_{\pi_{*}{X}} \omega - dH]_{\tg_{(n)}} = 0
\ee
And the only condition $\tg(n)$ needs to satisfy is
\be
T_{p}\tg_{n} = ker(\omega), p \epsilon \tg(n)
\ee
This characterizes the arbitrariness in the foliation. We shall now se its implementation for completely constrained systems. We introduce some notation that we shall use throughout the paper. Let us, for simplicity, consider a completely constrained first-class Hamiltonian system.
\be
H = v^{i}\phi_{i} = 0
\ee
where the $v_{i}$ are arbitrary Lagrange multipliers and $\phi_{i}$ are the constraints.\\
The equation
\be
\frac{\delta H}{\delta v_{a}} = 0
\ee
gives us the constraint
\be
\phi_{a}(q^{a}, p_{a}) = 0
\ee
which defines $\tg$. We now choose a local coordinate patch and define
\be
\omega = dp_{a} \wedge dq^{a}
\ee
The kernel of this will be the Hamiltonian vector fields whose integral surfaces define the motion of the system entirely. Thus the aim of using the presymplectic equation is to obtain the Hamiltonian vector field. And so
\be
[\iota_{\pi_{*}{X}} \omega - dH]_{\tg_{(n)}} = 0
\ee
gives us $\pi_{*}X = X_{H}$. now, when we take into account the vanishing of the constraints on the sub manifold, we attain the actual geometric form of the presymplectic equation
\be \left(X_{H}\right)^{\flat}|_{\tg_{(n)}}=0 \ee
(henceforth just $\tg$ will denote the final constraint sub manifold)
We now define the flow of a function along these integral surfaces as the map
\be
f^{\lambda}_{H}[g(q,p)] = g(q,p;\lambda)
\ee
where $\lambda$ is the parameter arising from the parametrization of the flow of$X_{H}$ which is again, completely arbitrary.The flow is determined by solving the Cauchy problem
\bqn
f^{0}_{H} &=& g \n \\
\frac{d}{d\lambda} f^{\lambda}_{H}[g] &=& X_{H}[g]
\eqn
which is solved by
\be
f^{\lambda}_{H}[g] = \sum_{n = 0}^\infty \frac{\lambda^{n}}{n!} X^{n}_{H}[g]
\ee
Now Hamilton's equatons are defined as the evolution of the flow of

\be
z^{a} = {q^{i}\choose p_{i}}
\ee
Therefore,
\be
\frac{d}{d\lambda} f^{\lambda}_{H}[g] = X_{H}[z^{a}]
\ee
giving,
\be
f^{0}_{H}[z^{a}] = {q^{i}\choose p_{i}}
\ee
And so the solution of Hamilton's equations is
\be
f^{\lambda}_{H}[z^{a}] = \sum_{n = 0}^\infty \frac{\lambda^{n}}{n!} X^{n}_{H}{q^{i}\choose p_{i}}
\ee
\sss{Relation to the Hamilton-Jacobi theory}
Consider, on $\tg$, the symplectic form 
\be
\pi^{*}\omega = dp \wedge dq = \pi^{*}d\theta = d(\pi^{*}\theta)
\ee
 If we choose a local coordinate patch $(q, p)$ on $\Gamma$ for which
\bqn
\theta = p dq \n \\
d\theta = \omega = dp \wedge dq
\eqn
Assuming that the phase space is simply connected,
\bqn
d(\pi^{*}\theta - \theta) &=& 0 \n \\
\pi^{*}\theta - \theta &=& dS
\eqn
This gives
\bqn
\frac{\p S}{\p Q_{i}} &=& P_{i} \n \\
\frac{\p S}{\p q_{i}} &=& p_{i}
\eqn
Assuming the Hamiltonian constraint has the form
\be
(g^{ab}p_{a}p_{b} + V(q_{a})) - E= H
\ee
$\frac{\delta H}{\delta v_{a}} = 0$ gives
\be
g^{ab}p_{a}p_{b} + V(q_{a}) = E
\ee
and from the above
\be
g^{ab}\frac{\delta S}{\delta q^{a}}\frac{\delta S}{\delta q^{b}} + V(q^{a}) = E
\ee
which is the Hamilton-Jacobi equation.
\subsection{Relational particle dynamics}
In this section relational particle dynamics, where we basically try and eliminate all the redundancy present in many particle dynamical theories. We begin by considering the configuration space for an n-particle dynamical system moving in a 3 dimensional space:
\be \mathcal{C}=\mathbb{R}^{3n} \ee
now this configuration space has redundency by virtue of translational, and Rotational invariance, which constitute the actions of the Euclidean group $Euc(3)$ which will have 6 dimensions (3 rotations + 3 translations), so we shall consider the Relative Configuration space:
\be \mathcal{C}_{rel}=\mathbb{R}^{3n}/Euc(3)\ee
Is this sufficient? - if we were considering a a universe consisting of but 3 particles, such that the only meaningful notion of position is the separation between the particles , we sould find that for the size of the triangles or straight lines formed by these particles the area and length should have no meaning whatsoever, this is due to the fact that a truly relational system isn't embedded in any background space, and so scale too must be eliminated, thus in totality we have to factor out scale as well, leaving us with a 3N-7 dimensional "shape space". The phase space will be the manifold $T^{*}\mathcal{C}\ni(q^{a},p_{a})$ and $$\tilde{q}^{a}=b(q^{a}+c^{a}-\epsilon^{abc}d_{b}q_{c})$$ is the "corrected" position. The b,c and d are the auxialry gauge variables corresponding to dilations (scale changes), translations and rotations respectively. Now, the Hamiltonian for this theory is given by
\bqn \mathcal{H}=N(\sum_{a} \frac{p_{a}p^{a}}{2m_{a}}+b^{2}U(\tilde{q}))-E) \\ +v_{1}(\sum \delta^{ab}p_{a}q_{b})+v_{2}(\sum p_{a})
 +v^{a}_{3}(\sum \epsilon^{abc}q_{b}p_{c})\eqn
Firstly, its evident that this theory is an example of a completely constrained Hamiltonian system having only first class constraints, secondly we find that the auxiliary gauge variables in the corrected position are actually the velocities with respect to the lapses of each of the parts of the Hamiltonian. This Hamiltonian is of the form
$$\mathcal{H}=NH+\sum v_{a}\phi_{a}$$
Now, we can attain the Presymplectic equation by the following steps: first, we find that
\bqn \frac{\delta \mathcal{H}}{\delta N}=0=>\sum_{a} \frac{p_{a}p^{a}}{2m_{a}}+b^{2}U(\tilde{q}))=E\\
 \frac{\delta \mathcal{H}}{\delta v_{a}}=0=>D=\sum \delta^{ab}p_{a}q_{b}=0;\\
T^{a}=\sum p^{a}=0;\\
J^{a}=\sum \epsilon^{abc}q_{b}p_{c}=0;
\eqn
The above constraints are known as the "Jacobi Constraints". Now the Jacobi constraints define $\tg$ and we find that $$\tg=T^{*}\mathbb{R}^{3n}/Euc(3)\times \mathcal{D}$$ where $\mathbb{R}^{3n}/Euc(3)\times \mathcal{D}$ is Shape space ($\mathcal{D}$ is the group of dilations). We can move forward and attain the Presymplectic equation.
\be [\iota_{X_{\mathcal{H}}}(dp_{a}\wedge dq^{a})-d\mathcal{H}]|_{\tg}=0 \ee
which gives us:
\bqn X_{\mathcal{H}}=(N\frac{p_{a}}{m_{a}}+v_{2}+v_{1}q_{a}+\\ \epsilon^{abc}v^{a}_{3}q_{b})\frac{\partial}{\partial q^{a}}\\
-(N b^{2}\frac{\partial U}{\partial q^{a}}-v_{1}q_{a}+\epsilon^{acb}v^{a}_{3}p_{c})\frac{\partial}{\partial p_{a}}\eqn
which satisfies:
\be \left(X_{\mathcal{H}}\right)^{\flat}|_{\tg}=0 \ee
Now we can identify what part of the Hamiltonian constraint is responsible for physical evolution:
\be \mathcal{E}_{NH}=(N\frac{p_{a}}{m_{a}})\frac{\partial}{\partial q^{a}}-(N b^{2}\frac{\partial U}{\partial q^{a}})\frac{\partial}{\partial p_{a}} \ee
(i.e. the vector field generated by the Hamiltonian constraint). It isn't hard to see that $$p^{a}=m\mathcal{E}_{NH}[q^{a}]$$
and so, by plugging this back into $\frac{\delta H}{\delta N}=0$ we find that: \be N=\sqrt{\frac{m\mathcal{E}_{NH}[q_{a}]\mathcal{E}_{NH}[q^{a}]}{2(E-U)}} \ee Thus we have re-derived the Ephemeris-Lapse correspondence.
\subsection{The Relativistic Free Particle}
Now the above formalism will be applied to the relativistic free particle traveling through Riemannian (or Pseudo Riemannian) spacetime. Note that this chapter is presented purely to exhibit some of the features of presymplectic dynamics, but not as a gravitational theory, for which we prescribe to the three-space approach (in the next paper). As of now, we shall consider the Hamiltonian
\be \mathcal{H}=N(g^{\mu \nu}p_{\mu}p_{\nu}-m^{2}) \ee
and we arrive at $\tg$ via:
$$\frac{\delta \mathcal{H}}{\delta N}=0=>g^{\mu \nu}p_{\mu}p_{\nu}=m^{2}$$ 
where $g^{\mu \nu}$ is a Riemannian (or Lorentzian) metric.
This defines a mass-hyperboloid $\mathcal{K}_{m}$ and we identify this with $\tg \ni (p_{\mu}, q^{\mu})$. As we did in the previous section, we attain the  Hamiltonian vector field tautologically via
\be [\iota_{X_{\mathcal{H}}}(dp_{\mu}\wedge dq^{\mu})-d\mathcal{H}]|_{\tg}=0 \ee
and in this case we get
\be X_{\mathcal{H}}=Np^{\mu}\frac{\partial}{\partial q^{\mu}}-N\frac{\partial g^{\mu \nu}}{\partial q^{\mu}}p_{\mu}p_{\nu}\frac{\partial}{\partial p_{\mu}} \ee
and it satisfies
\be \left(X_{\mathcal{H}}\right)^{\flat}|_{\mathcal{K}_{m}}=0 \ee. By the very same calculation which we did for the previous system, we can find the Lapse, and here it turns out to be
\be N=\sqrt{\frac{g^{\mu \nu}v_{\mu}v_{\nu}}{m^{2}}} \ee.
where $v_{\mu}=X_{\mathcal{H}}[q_{\mu}]$.

\subsection{Geodesics and Hamiltonian Flows}
In the previous section, the Hamiltonian vector field satisfying the presymplectic equation for a free relativistic particle was obtained now we shall attain the solutions via the flow of this vector field. In particular let us condier the flow of the position with respect to this Hamiltonian vector field, i.e
\be f^{\lambda}_{\mathcal{H}}[q^{\mu}]=\sum_{n=0}^{n=\infty}\frac{1}{n!}X^{n}_{\mathcal{H}}[q^{\mu}] \ee
from this, we know that
\be \frac{\partial^{2} f^{\lambda}_{\mathcal{H}}[q^{\mu}]}{\partial \lambda^{2}}=\frac{1}{2}X^{2}_{\mathcal{H}}[q^{\mu}]\ee
now $X_{\mathcal{H}}[q^{\mu}]=p^{\mu}/N$ so, $$X^{2}_{\mathcal{H}}[q^{\mu}]=\frac{1}{N}X_{\mathcal{H}}[p^{\mu}]=-N\frac{\partial g^{\mu \nu}}{\partial q^{\mu}}v_{\mu}v_{\nu}$$\\
By the definition of a flow$f^{\lambda}_{X_{\mathcal{H}}}[q^{\mu}]=q^{\mu}(\lambda)$, and putting all of this back into equation (8.50), and noting that $g^{\mu \nu}v_{\mu} v_{\nu}=g_{\mu \nu}v^{\mu}v^{\nu}$ we get
\be \frac{\partial^{2}q^{\xi}}{\partial \lambda^{2}}=-\frac{N}{2}\Gamma^{\xi}_{\mu \nu}v^{\mu}v^{\nu} \ee
The above equation (for on shell particles) is the geodesic equation. This shows us that the Hamiltonian flows on phase space and the geodesics on configuration space physically describe the very same dynamics of the system.
\s{Concluding remarks}
In this paper it has has been shown that the presymplectic equation is  adequate to describe the dynamics of reparameterization invariant, completely constrained Hamiltonian systems. In all the systems considered thus far, the constraints saturate the theories, and even thought they help define the specific domain of their dynamics on phase space i.e. the constraint submanifold via the presymplectic algorithm,  they and so the total Hamiltonian, do not explicitly feature in the final equation.
The Hamiltonian vector field is only derived from the natural tautology.
Also, the novelty of our formalism lies in the adoption of the interpretation of Barbour et. al. i.e. Machian Temporal Relationalism and its incorporation into a canonical phase space- framework, which can also account for the Ephemeris time and the violation of Dirac's theorem with regard to first class Hamiltonian constraints. More shall be said about the Barbour interpretation in our next paper in the context of canonical General Relativity.

\begin{acknowledgments}
This work was carried out at the \tit{Centre for Fundamental Research and Creative Education}, Bangalore, India. We would like to acknowledge the Director Ms. Pratiti B R for facilitating an atmosphere of free scientific inquiry so conducive to creativity. We would also like to thank our fellow researchers Magnona H Shastry, Madhavan Venkatesh, Karthik T Vasu and Arvind Dudi.
\end{acknowledgments}

\np
\printindex

\end{document}